\documentclass[published]{JHEP3} 

\JHEP{00(2010)000}

\JHEPspecialurl{http://jhep.sissa.it/JOURNAL/JHEP3.tar.gz}

\usepackage{epsfig,multicol}

\newcommand\fverb{\setbox\pippobox=\hbox\bgroup\verb}
\newcommand\fverbdo{\egroup\medskip\noindent%
            \fbox{\unhbox\pippobox}\ }
\newcommand\fverbit{\egroup\item[\fbox{\unhbox\pippobox}]}
\newbox\pippobox
\title{The Stokes Phenomenon and Quantum Tunneling for de Sitter Radiation in Nonstationary Coordinates}

\author{Sang Pyo Kim\\
 Department of Physics, Kunsan National University, Kunsan 573-701,
 Korea\\Asia Pacific Center for Theoretical Physics, Pohang 790-784,
 Korea\\ E-mail: \email{sangkim@kunsan.ac.kr}}

\received{}        
\revised{}
\accepted{}        

\preprint{\hepth{0710.0915}}  

\abstract{We study quantum tunneling for the de Sitter radiation in the planar
coordinates and global coordinates, which are nonstationary
coordinates and describe the expanding geometry. Using the phase-integral approximation
for the Hamilton-Jacobi action in the complex plane of time, we obtain the
particle-production rate in both coordinates and derive the additional sinusoidal factor
depending on the dimensionality of spacetime and
the quantum number for spherical harmonics in the global coordinates.
This approach resolves the factor of two problem in the
tunneling method.}


\begin{document}

\section{Introduction}

Black hole physics has been a key issue in gravity since Hawking's
discovery of black hole radiation \cite{Hawking}. Black holes radiate
thermal spectra with the temperature determined
by the surface gravity at the event horizon. One method to get
the Hawking radiation is the Bogoliubov transformation
between the in-falling state and the out-going state scattered by
the event horizon \cite{Hawking}. Recently Parikh and Wilczek have suggested an
intuitive interpretation of the Hawking radiation as quantum
tunneling through the event horizon of particles produced
from vacuum fluctuations near the horizon \cite{Parikh-Wilzcek}.
The tunneling interpretation has
revitalized the study of Hawking radiation in various black holes.
Other phenomena for particle
production are the Schwinger mechanism \cite{Schwinger}
and the expanding geometry \cite{Parker}.
The Schwinger mechanism can be understood as
quantum tunneling process of virtual pairs from the Dirac sea due to
the electric field.

The tunneling method has been mostly applied to static or stationary
black holes, where the event horizon provides a causal boundary for
tunneling of particles from the interior to the exterior.
Then a question may be raised whether the tunneling process is unique to
the static or stationary black holes. The de Sitter spacetime, describing
the geometry of an expanding spacetime, has a few
coordinate systems, one of which is the static coordinates that
exhibit the cosmological event
horizon with the Gibbons-Hawking temperature \cite{Gibbons-Hawking}.
The tunneling method has been applied to the
static coordinates of de Sitter radiation \cite{Padmanabhan02,Parikh02,Medved02,Zhang05,Kim08}.
In the static coordinates of the de Sitter spacetime in four dimensions
\cite{Spradlin-Strominger-Volovich,Moss}
\begin{eqnarray}
ds^2 = -(1 - \frac{r^2}{r_H^2})dt^2 + \frac{dr^2}{1 - \frac{r^2}{r_H^2}}
+ r^2 d\Omega_2^2, \label{static coord}
\end{eqnarray}
the cosmological event horizon is located at $r_{dS} = r_H$ and has the
Gibbons-Hawking temperature $T_{dS} = 1/(2 \pi r_H)$. Volovik used
the fluid metric for de Sitter spacetime, which is still a stationary
coordinates system, and derived the de Sitter radiation via quantum tunneling \cite{Volovik}.

In this paper we study the particle emission via
tunneling in nonstationary coordinates for de Sitter spacetimes.
Since Parker has shown that expanding spacetimes also
create particles \cite{Parker},  the particle-production rate in de Sitter spacetimes
has been known for many years \cite{Mottola,Bousso-Maloney-Strominger}.
The WKB approximation in complex time has been employed for the de Sitter radiation \cite{Biswas}.
In the Schwinger mechanism charged particle
pairs can also be produced by time-dependent electric fields or a constant electric field
in time-dependent gauge. Hence it is worthy to investigate the de Sitter radiation via tunneling in 
the planar coordinates and the global coordinates, these being
nonstationary coordinates. Other motivation is the derivation of
the Friedmann equation for de Sitter spacetime from
the first law of black hole thermodynamics \cite{Cai-Kim}.

In expanding spacetimes or time-dependent electric fields,
the ingoing positive frequency solution of a quantum field splits into
the outgoing positive and negative frequency solutions. The ratio of
the outgoing negative solution to the ingoing positive solution determines
the mean number of produced particles \cite{Parker}. The mean number is
approximately given by the imaginary part of the Hamilton-Jacobi action, which in turn
is determined by a contour integral in the complex plane of time \cite{Kim-Page07}. This
approach based on the phase-integral approximation \cite{Froman-Froman}
generalizes the tunneling idea to time-dependent electric fields \cite{Kim-Page07}.
Other field theoretical methods are also attempted
in time-dependent contexts \cite{Vakkuri-Kraus,Srinivasan,Shankaranarayanan02}.

We apply the phase-integral approximation to
the planar and global coordinates of de Sitter spacetime and obtain
the de Sitter radiation. The phase-integral approximation for
quantum tunneling recovers not only the Boltzmann factor
for the particle-production rate but also from the Stokes phenomenon \cite{Dumlu-Dunne} the sinusoidal factor
that depends on the dimensionality of spacetime and the quantum number for spherical harmonics.
One controversial issue in the tunneling method is the factor of two problem, according to which some coordinates give the temperature for radiation twice as big as the Hawking temperature \cite{Chowdhury,Akhmedov,Akhmedov07,Nakamura}.
We show that the tunneling method in nonstationary coordinates properly yields
the Hawking temperature and thus resolves the factor of two problem.

The organization of this paper is as follows. In section 2 we
briefly review the Schwinger mechanism in the time-dependent gauge
and formulate the pair-production rate via quantum tunneling in the complex time.
In section 3 we derive the de Sitter radiation via quantum tunneling in the planar
and global coordinates and explain the origin of sinusoidal factor that depends on
dimensionality of spacetime and quantum number for spherical harmonics.
Finally, we summarize the tunneling method in nonstationary
spacetimes and discuss the problem of the factor of two in section 4.

\section{Schwinger Mechanism in Time-Dependent Gauge}

The quantum electrodynamics (QED) analog for static or stationary black holes is the Coulomb gauge,
in which the field equation becomes the tunneling problem. The analogy between the
Schwinger mechanism and the Hawking radiation has been studied in Refs. \cite{Stephens,Parentani97,Kim07}.
In QED the pair-production rate can be found in various
methods, such as the phase-integral approximation
\cite{Kim-Page02,Kim-Page06,Kim-Page07} and the worldline instanton
method \cite{Dunne-Schubert,DWGS}. The constant electric field has not only the Coulomb gauge but also the
time-dependent gauge, in which the field equation becomes either tunneling under the barrier
or transmission over the barrier.

We briefly review the phase-integral approximation for pair production
in the time-dependent gauge \cite{Kim-Page07}, which is the QED analog for nonstationary spacetimes.
The component of a charged scalar field in the gauge potential $A_{\parallel}(t)$
satisfies the equation
\begin{eqnarray}
\ddot{\phi}_{\bf k} (t) + Q_{\bf k} (t) \phi_{\bf k} (t)= 0,
\end{eqnarray}
where
\begin{eqnarray}
Q_{\bf k} (t) = m^2 + {\bf k}_{\perp}^2 + (k_{\parallel}+qA_{\parallel} (t))^2.
\end{eqnarray}
Here $m$ and $q$ are the mass and charge of the particle, and ${\bf k}_{\perp}$ and $k_{\parallel}$ are
the momentum component transverse and parallel to the electric field, respectively.
The magnitude square of the ratio of the outgoing negative frequency solution to the ingoing
positive solution is the mean number of produced pairs. In Ref.
\cite{Kim-Page07}, the WKB instanton action for the Hamilton-Jacobi equation
\begin{eqnarray}
2 {\rm Im} S_{\bf k} = i \oint \sqrt{Q_{\bf k} (t)} dt,
\end{eqnarray}
where the contour integral is taken outside of a loop enclosing two
roots of $Q_{\bf k} (t)$ in the complex plane of time,
approximately gives the mean number of produced pairs
\begin{eqnarray}
\bar{n}_{\bf k} = e^{- 2 {\rm Im} S_{\bf k} }. \label{mean num}
\end{eqnarray}
The action $2 {\rm Im} S_{\bf k}$ takes the probability into account, which is equivalent to the detailed balance of the emission rate to the absorption rate \cite{Stephens,Srinivasan}.

The instability of the vacuum due to pair production is characterized by the vacuum persistence, which is related to the mean number of produced pairs.
The vacuum persistence for bosons is the probability for the in-vacuum to remain
in the out-vacuum \cite{Kim-Lee-Yoon08,Hwang-Kim}
\begin{eqnarray}
\vert \langle {\rm out } \vert {\rm in} \rangle \vert^2 =
e^{ - VT \sum_{\bf k} \ln (1 + \bar{n}_{\bf k})},
\end{eqnarray}
where $V$ is the volume and $T$ is the duration.
For small pair production $\bar{n}_{\bf k} \ll 1$, the decay rate per unit volume
and per unit time is approximately given by the total mean number of produced pairs:
\begin{eqnarray}
\frac{\Gamma}{VT} = 1 - \vert \langle {\rm out } \vert {\rm in} \rangle \vert^2 \approx \sum_{\bf k} \bar{n}_{\bf k}.
\end{eqnarray}
In the phase-integral approximation for tunneling, it is understood that the emission rate is
determined by Eq. (\ref{mean num}).

\section{de Sitter Spacetimes}

The de Sitter spacetime has a few different coordinate systems. The (d+1)-dimensional de
Sitter spacetime has the planar
coordinates,\footnote{The planar coordinates are also called the
oblique coordinates and the global coordinates are the horizontal
coordinates \cite{Moss}.}
\begin{eqnarray}
ds^2 = -dt^2 + e^{2Ht} d{\bf x}^2_d,
\end{eqnarray}
where the units $c = \hbar =1$ are used, and also has the global coordinates
\begin{eqnarray}
ds^2 = -dt^2 + \frac{1}{H^2} \cosh^2 (Ht) d \Omega^2_d.
\end{eqnarray}
The field equation for a scalar field contains the d-dimensional Laplace operator and
harmonics \cite{Rubin-Ordonez}
\begin{eqnarray}
\nabla_d^2 u_{\kappa}({\bf x}) = - \kappa^2 u_{\kappa} ({\bf x}),
\end{eqnarray}
where for the planar coordinates $\nabla_d^2$ acts on
the d-dimensional Euclidean space and has
$\kappa^2 = {\bf k}^2$ while for the global coordinates $\nabla_d^2$ acts
on $S_d$ and has $\kappa^2 = l(l+ d-1)$, $(l = 0, 1, \cdots)$. Then the harmonic-components
of the scalar field, $\Phi (t, {\bf x}) = a^{-d/2} \sum_{\kappa} u_{\kappa}({\bf x}) \phi_{\kappa}$, satisfy
\begin{eqnarray}
\ddot{\phi}_{\kappa} (t) + Q_{\kappa} (t) \phi_{\kappa} (t) = 0, \label{com eq}
\end{eqnarray}
where
\begin{eqnarray}
Q_{\kappa} (t) = m^2 + \frac{\kappa^2}{a^2} - \frac{d}{2} (\frac{d}{2} -1)
\Bigl( \frac{\dot{a}}{a} \Bigr)^2 - \frac{d}{2} \frac{\ddot{a}}{a}.
\end{eqnarray}
Here $a(t) = e^{Ht}$  for the planar coordinates and $a(t) = \cosh(Ht)/H$ for global coordinates.
We now apply the phase-integral approximation and find the solution of the form
\begin{eqnarray}
\ddot{\phi}_{\kappa} (t) = e^{- iS_{\kappa} (t)},
\end{eqnarray}
in terms of the Hamilton-Jacobi action
\begin{eqnarray}
 S_{\kappa} (t) =  \int \sqrt{Q_{\kappa} (t)} dt. \label{HJ act}
\end{eqnarray}
The imaginary part of $S_{\kappa}$ is
responsible for the decay rate and gives the
Boltzmann factor for the de Sitter radiation,
\begin{eqnarray}
\Gamma_{\kappa} = |\phi_{\kappa}|^2 = e^{- 2 {\rm Im} S_{\kappa}}.
\end{eqnarray}

First, in the planar coordinates, Eq. (\ref{HJ act}) takes the form
\begin{eqnarray}
 S_{\bf k} (t) =  \int \sqrt{ \gamma^2 + {\bf k}^2 e^{-2Ht}} dt, \quad \gamma = \sqrt{m^2 - \frac{(dH)^2}{4}}.
\end{eqnarray}
Changing the variable $\tau = e^{Ht}$ and integrating from a turning point $\tau_- = - i k/\gamma$  to another $\tau_+ =  i k/\gamma$, we obtain a pure imaginary action
\begin{eqnarray}
S_{\bf k} = \frac{1}{H} \int_{- i \frac{k}{\gamma}}^{i \frac{k}{\gamma}} \frac{d \tau}{\tau^2}
\sqrt{{\bf k}^2 + \gamma^2 \tau^2} = i  \pi \frac{\gamma}{H}.
\end{eqnarray}
The analogy with the Schwinger mechanism suggests another way to find the instanton action.
The production rate of particles may be given by a contour integral \cite{Kim-Page07}
\begin{eqnarray}
2 {\rm Im} S_{\bf k} = i \oint \sqrt{\gamma^2 + {\bf k}^2
e^{-2Ht} } dt, \label{con int}
\end{eqnarray}
in the complex plane of time $t$. Expanding for large $\tau$ and
choosing a contour exterior to two branch cuts $\tau_{\pm}$, the contour integral
\begin{eqnarray}
2 {\rm Im} S_{\bf k} = i \frac{\gamma}{H} \oint
\sqrt{1 + \frac{{\bf k}^2}{(\gamma \tau)^2} } \frac{d \tau}{\tau}, \label{con int2}
\end{eqnarray}
 with the negative sign for residues, leads to the instanton action
\begin{eqnarray}
2 {\rm Im} S_{\bf k} = 2 \pi \frac{\gamma}{H}. \label{imag}
\end{eqnarray}

Second, in the global coordinates, the instanton action of the
Hamilton-Jacobi equation is given by
\begin{eqnarray}
S_{l} = \int \sqrt{\gamma^2 + \frac{(\lambda
H)^2}{\cosh^2(Ht)}} dt, \quad \lambda^2 = l(l+d-1) + \frac{d}{2}
\Bigl(\frac{d}{2} -1 \Bigr). \label{g-con int}
\end{eqnarray}
There are four turning points,
\begin{eqnarray}
e^{H t_{(a) \pm} } &=& - i (\lambda H/\gamma) \pm i \sqrt{(\lambda H/\gamma)^2 + 1}, \nonumber\\
e^{H t_{(b) \pm}} &=& + i (\lambda H/\gamma) \pm i \sqrt{(\lambda H/\gamma)^2 + 1}.
\end{eqnarray}
These are grouped into complex conjugate pairs $\{ t_{(a)+}, t_{(b)-} \}$ and $\{ t_{(a)-}, t_{(b)+} \}$.
Changing the variable to a conformal time, $\sinh(Ht) = \tan (\tau)$, the Hamilton-Jacobi action takes the
form
\begin{eqnarray}
S_{l} (t_{(a)}, t_{(b)}) = \frac{\gamma}{H} \sqrt{1 + \frac{(\lambda H)^2}{\gamma^2}} \int_{t_{(a)}}^{t_{(b)}}
\sqrt{1 - \frac{\frac{(\lambda H)^2}{\gamma^2} \sin^2 (\tau)}{1+ \frac{(\lambda H)^2}{\gamma^2}} } \frac{d \tau}{\cos(\tau)}. \label{g-con int2}
\end{eqnarray}
The integral can be done \cite{PBM}
\begin{eqnarray}
S_{l} (t_{(a)}, t_{(b)}) = i \pi \frac{\gamma}{H} + \pi \lambda. \label{g-con int3}
\end{eqnarray}
Each complex conjugate pair contributes $2 {\rm Im} S_{l} = 2 \pi \gamma/H$ and the crossing integral, for instance, from $t_{(a)-}$ to $t_{(b)-}$ yields the real part ${\rm Re} S_{l} = \pi \lambda$. 
Therefore, the Stokes phenomenon, Eq. (7) of
Ref. \cite{Dumlu-Dunne} (see also Ref. \cite{Froman-Froman}), gives the particle-production rate
\begin{eqnarray}
\bar{n}_{l} &\approx&  e^{- 2 {\rm Im} S_{l} (t_{(a)+}, t_{(b)-})}+ e^{- 2 {\rm Im} S_{l} (t_{(a)-}, t_{(b)+})}
\nonumber\\&& + 2 \cos \bigl(2 {\rm Re} S_{l} (t_{(a)-}, t_{(b)-}) \bigr) e^{- {\rm Im} S_{l} (t_{(a)+}, t_{(b)-}) - {\rm Im} S_{l} (t_{(a)-}, t_{(b)+})}.
\end{eqnarray}
In the asymptotic limit of large action ($S_{l} \gg 1$ and
$l \gg 1$), we approximate $\lambda \approx l + (d-1)/2$ and obtain
\begin{eqnarray}
\bar{n}_{l} \approx 4 \sin^2 \Bigl(\pi(l+ \frac{d}{2}) \Bigr) e^{- 2 \pi \frac{\gamma}{H}},
\end{eqnarray}
which is the leading term of particle-production rate in the (d+1)-dimensional de Sitter spacetime
\begin{eqnarray}
\bar{n}_l = \frac{\sin^2(\pi(l+ \frac{d}{2}))}{\sinh^2 (\frac{\pi \gamma}{H})}.
\end{eqnarray}
The sinusoidal factor is a consequence of the substructure of the Stokes phenomenon 
and explains the absence of particle production in odd dimensions \cite{Bousso-Maloney-Strominger,JMP}. In the second approach, changing the variable $u = \cosh (Ht)/H$
and expanding for large $u$, the contour integral leads to
\begin{eqnarray}
2 {\rm Im} S_{l} = i \frac{\gamma}{H} \oint \frac{d \tau}{\tau}
\frac{\sqrt{1 + \frac{\lambda^2}{(\gamma \tau)^2 } }}{\sqrt{1-
\frac{1}{(H \tau)^2}}} = 2 \pi \frac{\gamma}{H}.
\end{eqnarray}

In summary, we applied the phase-integral approximation to the planar coordinates
and the global coordinates of de Sitter spacetime
and obtained the Boltzmann factor for the mean number of produced particles
\begin{eqnarray}
\bar{n}_{\kappa} = e^{- 2\pi \frac{\gamma}{H}}.
\end{eqnarray}
The de Sitter radiation has the Gibbons-Hawking temperature $T_{\rm d S} = H/2 \pi$.
In the global coordinates, in addition to the Boltzmann factor, there is a sinusoidal factor that depends on the spacetime dimensions and the quantum number for spherical harmonics. That particles are not produced in odd dimensional de Sitter spacetimes is a consequence of the Stokes phenomenon in the global coordinates.

\section{Conclusion}

In this paper we have extended the tunneling idea for particle production to nonstationary
coordinates of de Sitter spacetimes. The tunneling method, which provides a physical intuition
to the Hawking radiation of black holes, has been also applied to static coordinates
of de Sitter spacetimes. It is thus interesting to investigate the tunneling method in nonstationary spacetimes. For this purpose we have used the analogy of particle production by expanding spacetimes with the Schwinger mechanism by electric fields.

In QED a constant electric field has the Coulomb gauge or the time-dependent gauge. 
The Coulomb gauge corresponds to a
static black hole, while the time-dependent gauge
corresponds to a nonstationary spacetime. Further, the pair-production
rate for a given momentum ${\bf k}$
\begin{eqnarray}
\bar{n}_{\bf k} = e^{- 2{\rm Im} S_{\bf k}} \label{Boltzmann fac}
\end{eqnarray}
is given by the Hamilton-Jacobi action in the complex plane of space or time \cite{Kim-Page07}
\begin{eqnarray}
2 {\rm Im} S_{\bf k} = \mp i \oint \sqrt{Q_{\bf k} (z)} dz. \label{WKB inst}
\end{eqnarray}
Here the upper sign is for the Coulomb gauge and $Q_{\bf k} (z)$ is the kinematic momentum 
in the electric field while the lower sign is for the time-dependent gauge 
and  $Q_{\bf k} (z)$ is the kinematic energy. In this sense Eqs. (\ref{Boltzmann fac}) and (\ref{WKB inst}) 
provide a unified tunneling method for the Coulomb gauge and the time-dependent gauge.

We have shown that the tunneling method formulated by Eqs. (\ref{Boltzmann fac}) and (\ref{WKB inst}) 
also applies to the planar and global coordinates of de Sitter spacetimes. The tunneling method 
recovers the Boltzmann factor for de Sitter radiation as shown in section 3 and, to our surprise, it 
yields the sinusoidal factor depending on the dimensionality of de Sitter spacetime and 
the quantum number for spherical harmonics in the global coordinates. The absence of particle production in the global coordinates of odd dimensional de Sitter spacetimes is a consequence of the Stokes phenomenon.

One controversial issue in the tunneling method is the factor of two problem \cite{Chowdhury,Akhmedov,Akhmedov07,Nakamura}, 
where the temperature is twice as big as the Hawking temperature. To resolve this problem,
either the isotropic coordinates \cite{ANVZ,NVZ} or Rindler coordinates 
\cite{Kim07} have been used. Indeed, in these coordinates the tunneling method with the upper sign of
Eqs. (\ref{Boltzmann fac}) and (\ref{WKB inst}) yields correctly the de Sitter radiation. 
However, tunneling method does not have such an ambiguity 
in the planar and global coordinates as shown in section 3. 

\section*{Acknowledgments}
The author thanks Gerald Dunne for explaining the Stokes phenomenon. And he also appreciates 
the warm hospitality of Yukawa Institute for Theoretical Physics of Kyoto University, 
where this paper was completed.
This work was supported by the Korea Science and Engineering
Foundation (KOSEF) grant funded by the Korea government (MOST) (No.
F01-2007-000-10188-0).

\end{document}